\documentstyle[12pt]{article}

\thicklines                         
\def\a{\alpha}
\def\b{\beta}

\def\d{\delta}
\def\e{\epsilon}                
\def\f{\phi}                    
\def\g{\gamma}

\def\j{\psi}

\def\l{\lambda}
\def\m{\mu}
\def\n{\nu}

\def\s{\sigma}                  

\def\O{\Omega}

\def\cp{{\cal P}}

\def\cu{{\cal U}}

\def\cbo{{\,\raise-.15ex\Sc [\,}}                       
\def\dg{^\dagger}                                     
 
\def\sbra#1{\left\langle #1\right|}             
\def\sket#1{\left| #1\right\rangle}             
\def\svev#1{\left\langle #1\right\rangle}       

\def\ddt#1{{\buildrel {\hbox{\LARGE .\kern-2pt.}} \over {#1}}}

\def\beq{\begin{equation}}
\def\eeq{\end{equation}}
\def\bqry{\begin{eqnarray}}
\def\eqry{\end{eqnarray}}

\def\secteq#1{ \setcounter{equation}{0}
               \renewcommand{\theequation}{#1.\arabic{equation}} }

\def\beqn#1{ \renewcommand{\theequation}{#1} 
             \begin{eqnarray} }
\def\eeqn{ \renewcommand{\theequation}{\arabic{equation}}
           \end{eqnarray} }

\def\beqr#1{ \setcounter{equation}{#1} 
             \begin{eqnarray} }

\def\eeqr{\end{eqnarray}}
\def\NON{\nonumber\\}
\def\beqrabc#1{ \setcounter{equation}{0}
                \renewcommand{\theequation}{#1\alph{equation}} 
                \begin{eqnarray} }
\def\beqrn#1#2{ \setcounter{equation}{#2}
                \renewcommand{\theequation}{#1.\arabic{equation}} 
                \begin{eqnarray} }
\def\seeq#1{eq.~(\ref{#1})}
\def\seEq#1{Eq.~(\ref{#1})}
\def\seeqs#1{eqs.~(\ref{#1})}
\def\seEqs#1{Eqs.~(\ref{#1})}
\def\seneq#1{~(\ref{#1})}

\def\NPB#1{Nucl. Phys. {\bf B#1}}

\def\PLB#1{Phys. Lett. {\bf B#1}}
\def\PRD#1{Phys. Rev. {\bf D#1}}

\def\PRL#1{Phys. Rev. Lett. {\bf #1}}
\def\PRP#1{Phys. Rep. {\bf #1}}

\def\sstyle{\scriptstyle}

\def\rhs{\mbox{r.h.s.} }

\def\ie{\mbox{\it i.e.} }
\def\eg{\mbox{\it e.g.} }

\def\frac#1#2{ {\sstyle {#1\over #2} } } 
\def\det#1{{\rm det}\left(#1\right)}

\def\tr{{\rm tr}\,}

\def\half{{1\over 2}}
\def\Re{{\rm Re\,}}

\def\det{{\rm det\,}}

\def\TT{\tilde{T}}

\begin{document}
\hyphenation{fer-mio-nic per-tur-ba-tive}

July 1998 \hfill TAUP--2510--98  

\begin{center}
\vspace{15mm}
{\large\bf Reducing Chiral Symmetry Violations \\
in Lattice QCD with Domain-Wall Fermions}
\\[15mm]
Yigal Shamir
\\[5mm]
{\it School of Physics and Astronomy\\
Beverly and Raymond Sackler Faculty of Exact Sciences\\
Tel-Aviv University, Ramat~Aviv,~69978~ISRAEL}
\\[15mm]
{ABSTRACT}
\\[2mm]
\end{center}

\begin{quotation}
The inverse of the fermion matrix squared
is used to define a transfer matrix for domain-wall fermions.
When the domain-wall height $M$ is bigger than one, 
the transfer matrix is {\it complex}.
Slowly suppressed chiral symmetry violations may then arise from all 
eigenvalues of the transfer matrix which are located near the unit circle.
Using a variable lattice spacing for the fifth coordinate
we enforce the strict positivity of the transfer matrix for any $M$. 
We furthermore propose a modified pseudo-fermion action, 
aimed to decrease the density of close-to-unity eigenvalues
of the (positive) transfer matrix, at the price of
a small renormalization of the coupling constant.
We explain why these changes may reduce chiral symmetry violations
in lattice QCD simulations.
\end{quotation}

 

\newpage
\noindent {\large\bf 1.~~Introduction}
\vspace{3ex}
\secteq{1}

Numerical QCD simulations~[1-4] using domain-wall fermions~[5-8]
reveal small, but still significant, lattice-artefact violations of 
chiral symmetries for $N_s \sim 15$,
where $N_s$ is the extent of the lattice in the fifth direction.
It is important to improve our understanding of 
such anomalous effects, and, hopefully, to devise 
new domain-wall actions capable of better suppressing them.
Both at tree-level and in perturbation
theory the anomalous effects decrease exponentially with increasing $N_s$.
This applies in particular to the additive radiatively-induced 
quark mass~\cite{jansen,bndr,PT,BSW}. In this paper we will thus focus on 
non-perturbative anomalous contributions to 
the chiral Ward identities introduced in ref.~\cite{VF} and first 
studied numerically in ref.~\cite{BS}.

Free domain-wall fermions have a single chiral zero mode on each  
four-dimensional boundary of a five-dimensional lattice
when the five-dimensional mass term (or ``domain wall height") $M$ 
is in the range $0<M<2$. As one moves to the range $2<M<4$
this zero mode disappears, and four new zero modes of the opposite chirality
appear on each boundary. 
Thus, the range $0<M<2$ supports a single (approximately) massless quark, 
and the range $2<M<4$ supports four (approximately) massless quarks 
with ``flipped" chiralities. This pattern generalizes to higher values of $M$, 
until for $M>10$ there are no zero modes at all.

With suitable subtractions to ensure its finiteness
(to be discussed in detail in this paper)
the $N_s\to\infty$ limit of domain-wall fermions can be written down
compactly in the overlap formalism~\cite{NNold,NNnew}.
In this formalism the (subtracted) fermion partition function is expressed 
in terms of the overlap of a second-quantized ground state 
with a reference state
or, more generally, as the ground-state expectation value
of an operator representing the boundary conditions in the 
$s$-direction~\cite{VF}.

In the ``old" overlap formalism~\cite{NNold}, 
the second-quantized hamiltonian is (minus) the logarithm of the transfer
matrix that hops domain-wall fermions a single site in the $s$-direction.
This formalism has the same massless-quark spectrum as above. 
The ``new" overlap formalism~\cite{NNnew}
involves a different hamiltonian which arises in a continuous
$s$-coordinate limit. The new overlap
(and the related four-dimensional, non-local,
chirally invariant action~\cite{HN07})
has the same massless-quark spectrum for $0<M<2$. In the range $2<M<4$
four massless quarks with flipped chiralities appear
as before, but the original massless quark {\it remains} in the spectrum.
As $M$ is further increased more massless quarks appear,
alongside with all the previous ones that keep staying in the spectrum. 
This different free-field spectrum agrees with the pattern of level
crossing found in a smooth instanton background~\cite{SCRI1}. 

A more fundamental difference between the two overlap formulae
is found in the properties of the transfer matrix. 
In the new overlap formalism the transfer matrix 
is strictly positive for all values of $M$, simply because it is defined
from the outset as $\exp(-H)$, where $H$ is a well-defined
local lattice hamiltonian.
In the old overlap formalism, on the other hand,
the transfer matrix is positive only for $0<M<1$,
as in the framework of domain-wall fermions from which it is derived.
We argue below that for $M>1$ both positivity and hermiticity
of the domain wall's transfer matrix are lost, and its eigenvalues are 
in general {\it complex}.

The aim of this paper is to study what could be
the dominant anomalous contributions to chiral Ward identities,
and to suggest methods for suppressing them.
A key role will be played by a (first quantized) transfer matrix 
related to the inverse of the fermion matrix squared. 
For flavor non-singlet chiral symmetries, the anomalous term 
in the lattice Ward identities~\cite{VF} involves the correlations of
fermion operators on the $s=1$ and $s=N_s$ boundaries with 
fermion operators at $s \sim N_s/2$. Such correlations decrease
exponentially if the transfer matrix has no eigenvalue with an absolute value 
close to one. Conversely, if there are many such eigenvalues,
one expects a much slower (presumably power-law) decrease
of anomalous correlations.

Numerical results~\cite{BS,COL} suggest an optimal value of $M$
between 1.6 and 1.7 for $\b \sim 6$. In this range, the transfer
matrix of standard domain-wall fermions is complex.
Slowly decreasing anomalous contributions may then
arise from all eigenvalues of the transfer matrix which are
close to the unit circle.

The positivity of the transfer matrix can be enforced for any $M$
by restricting the range of $a_5$, the lattice spacing 
for the fifth coordinate~\cite{KY}, a result that
could have been anticipated in view of the relation between
the old and new overlaps. 
The transfer matrix is strictly positive for $Ma_5<1$.
(Thus, for $M<2$ it is sufficient to take \eg $a_5 = 0.5$.)
In this case, only close-to-unity eigenvalues~\cite{PV,VF,NNnew,SCRI2,ssrs}
may lead to significant anomalous contributions.  
We propose a method to reduce the density of such
eigenvalues in dynamical simulations. The method consists of a modification
of the pseudo-fermion (Pauli-Villars) part of the action,
and it might be effective already for modest values of $N_s$.
The main side-effect of the modified action is a small renormalization
of the coupling constant.

This paper is organized as follows. Domain-wall fermions with a variable 
lattice spacing for the fifth coordinate are reviewed
in Sec.~2. The (first quantized) transfer matrix is introduced
in Sec.~3. The case of a complex transfer matrix is discussed in Sec.~4.
The modified pseudo-fermion action is introduced and discussed 
in Sec.~5. Our conclusions are given in Sec.~6, 
and some technical details are relegated to two appendices.

\vspace{5ex}
\noindent {\large\bf 2.~~Domail-Wall Fermions with a variable $s$-spacing}
\vspace{3ex}
\secteq{2}

Allowing for a variable lattice spacing $a_5$ for the  
fifth coordinate~\cite{KY}, the domain-wall fermion matrix is
\beq
D_F = \left(\begin{array}{cccccccc}
a_5 D-1 & P_R & 0 & 0 & \ldots & 0 & 0 & -m a_5 P_L \\
P_L & a_5 D-1 & P_R & 0 & \ldots & 0 & 0 & 0 \\
0 & P_L & a_5 D-1 & P_R & \ldots & 0 & 0 & 0 \\
\vdots & \vdots & \vdots & \vdots & \ddots & \vdots & \vdots & \vdots \\
0 & 0 & 0 & 0 & \ldots & P_L & a_5 D-1 & P_R \\
-m a_5 P_R & 0 & 0 & 0 & \ldots & 0 & P_L & a_5 D-1
\end{array}\right) 
\label{dwf}
\eeq
The four-dimensional lattice spacing $a$ is set to unity.
The above matrix structure corresponds to the fifth coordinate $s$, which
we assume to take the values $s=1,2,\ldots,N_s$. 
Each entry is a four-dimensional matrix. $P_{R,L}={1\over 2}(1\pm\g_5)$
denote chiral projectors, and $D$ is the Wilson-Dirac operator 
\beq
\label{dpr}
   D = \left(\begin{array}{cc}
   M-W    & C     \\
   -C\dg  & M-W
       \end{array}\right)\,,
\eeq
where
\bqry
   C_{xy} & = & \half \sum_\m \left[\d_{x+\hat\m,y} U_{x\m}
                 - \d_{x-\hat\m,y} U^\dagger_{y\m} \right] \s_\m \\
   W_{xy} & = & 4\d_{xy} -\half \sum_\m \left[\d_{x+\hat\m,y} U_{x\m}
                 + \d_{x-\hat\m,y} U^\dagger_{y\m} \right] \,.
\eqry
The matrices $C$ and $W$ correspond to the kinetic and Wilson term
respectively. We also define
\beq
  B_{xy} = (1 - M a_5)\d_{xy} + a_5 W_{xy} \:.
\label{B}
\eeq
($B$ is proportional to the identity matrix in spinor space.
The latter may be either two-by-two or four-by-four;
the correct meaning can be inferred from the context.)

Assume momentarily a semi-infinite range for the $s$-coordinate.
For $\sin(p_\m)=0$, $\m=1,\ldots,4$, the free-field equation 
has a right-handed homogeneous solution 
\beq 
  \j_R^0(s;p_\m) = B_0^s(p_\m) = [1+a_5(W_0(p_\m)-M)]^s \:,
\eeq
where the subscript zero denotes free-field quantities,
$W_0(p_\m)=\sum_\m (1-\cos(p_\m))$, and 
at the corners of the Brillouin zone $W_0(p_\m)$ takes the values 
$0,2,\ldots,8$.
The above homogeneous solution is a zero mode (\ie it is normalizable) provided 
\beq
  -1 < 1+a_5(W_0(p_\m)-M) < 1 \:.
\label{zm}
\eeq
We depart from the original domain-wall framework by replacing 
the constraint on the lower bound with the stronger one 
\beq
  0 < 1 - M a_5 \:.
\label{Ma5}
\eeq
Since $W$ is a positive matrix, this implies
\beq
  0 < 1+a_5({\rm spec}\,(W)-M) \:.
\eeq
The last condition ensures the strict {\it positivity} of $B$.
The upper-bound constraint in\seneq{zm} is simply
\beq
  W_0(p_\m)-M<0 \,,
\eeq
which is evidently independent of $a_5$.
For $0<M<2$ there is a single zero mode at $p_\m=0$. 
If we increase $M$ while decreasing $a_5$ to maintain the constraint\seneq{Ma5},
new zero modes will appear at $M=2,4,6,8,$ while
all the zero modes from smaller values of $M$ will remain in the spectrum.

Assuming $0<M<2$, 
on a finite lattice the right-handed part of the quark field corresponds
to the above zero mode, whereas the left-handed part corresponds to
another zero mode located near the $s=N_s$ boundary.
Both at tree level and in perturbation theory, for $m=0$
there is a small mixing between the two chiral modes that vanishes 
exponentially with $N_s$ \cite{jansen,bndr,PT,PV}.
The parameter $m$ in \seeq{dwf} is related to the bare quark mass.
An easy way to see this is to invoke the free domain-wall hamiltonian.
For $m=0$, the eigenvalues $E(p_k)$ of helicity eigenstates are given by
$E^2=\sum_{k=1}^3 \sin^2(p_k)$. 
For $m\ne 0$ we find using first-order perturbation theory that
$E(0)=m_q$ where
\beq
  m_q = m M a_5 (2 - M a_5) \:,
\label{mquark}
\eeq 
generalizing the result of ref.~\cite{bndr}.

\vspace{5ex}
\noindent {\large\bf 3.~~Propagators and transfer matrices}
\vspace{3ex}
\secteq{3}

The familiar relation $D = \g_5 D^\dagger \g_5$, valid for
the Wilson-Dirac operator, generalizes in the case of domain-wall fermions to 
\beq
  D_F = {\cal R}\g_5 \, D_F^\dagger \, {\cal R}\g_5 \:,
\label{rg5}
\eeq
where ${\cal R}_{s s'} = \d_{s,N_s+1-s'}$~\cite{VF}. 
\seEq{rg5} implies relations between second-order operators:
$D_F^\dagger D_F = ({\cal R}\g_5 \, D_F)^2 = 
{\cal R}\g_5  (D_F D_F^\dagger) {\cal R}\g_5$. 
We also note that ${\cal R}\g_5 \, D_F$ is hermitian. 
For definiteness we will focus on the operator
$D_F D_F^\dagger$. One has
\beq
D_F D_F^\dagger = \g_5 B^{1/2}\, \O\, \g_5 B^{1/2} \:,
\label{DDag}
\eeq
where explicitly
\beq
\O = \left(\begin{array}{cccccccc}
X_{++} & -1 & 0 & 0 & \ldots & 0 & 0 & X_{-+} \\
-1 & Y & -1 & 0 & \ldots & 0 & 0 & 0 \\
0 & -1 & Y & -1 & \ldots & 0 & 0 & 0 \\
\vdots & \vdots & \vdots & \vdots & \ddots & \vdots & \vdots & \vdots \\
0 & 0 & 0 & 0 & \ldots & -1 & Y & -1 \\
X_{+-} & 0 & 0 & 0 & \ldots & 0 & -1 & X_{--}
\end{array}\right)
\label{Omega}
\eeq
and
\bqry
  Y & = & 2 + a_5^2 \g_5 B^{-1/2}\, D D^\dagger\, \g_5 B^{-1/2} 
\label{Y} \\
  X_{++} & = & Y + P_L B^{-1} ((m a_5)^2-1) \\
  X_{--} & = & Y + P_R B^{-1} ((m a_5)^2-1) \\
  X_{+-} & = & X_{-+} = m a_5 \,.
\eqry
The $B^{\pm{1\over 2}}$ factors have been introduced for later convenience.
Thanks to our insistence on the strict positivity of $B$, 
these factors are strictly positive too. 
An interesting observation is that
\beq
  Y = \TT + \TT^{-1} \,,
\label{YTT}
\eeq
where 
\beq
  \TT = 
  \left( \begin{array}{cc}
  B^{-1} + a_5^2 B^{-1/2}\, C C^\dagger\, B^{-1/2} 
  &  a_5 B^{-1/2}\, C\, B^{1/2}    \\
  a_5 B^{1/2}\, C^\dagger\, B^{-1/2} &  B
       \end{array}\right) \,.
\label{TT}
\eeq
The basic properties of the {\it transfer matrix} $\TT$ follow from
\beq
  \TT=K^\dagger K \,,
\label{TKK}
\eeq
where 
\beq
   K = \left( \begin{array}{cc}
              B^{-1/2}      &  0       \\
              a_5 C^\dagger\, B^{-1/2} &  B^{1/2}
       \end{array}\right) \,,
\qquad
   K^\dagger = \left( \begin{array}{cc}
              B^{-1/2}  &  a_5 B^{-1/2}\, C     \\
              0  &  B^{1/2}
       \end{array}\right) \,,
\label{K}
\eeq
which imply that $\TT$ is strictly positive, bounded, and has 
$\det\TT=1$.

The first-quantized transfer matrix usually encountered in the context 
of domain-wall fermions is $T=K K^\dagger$. Evidently, $T$ and $\TT$ have
the same spectrum, and (up to normalization) the eigenvectors of $T$ are 
obtained from those of $\TT$ by multiplication with $K$.
The appearance of $\TT$ instead of $T$ is due to a technical reason 
that we explain later.

\seEqs{Y} and\seneq{YTT} imply that an eigenvalue of $\TT$ 
is equal to one {\it iff} the hermitian
operator $\g_5 D$ has a zero mode. The last condition implies
$\det(D)=0$, an equation which defines a measure-zero subset 
of the gauge field configuration space. In the rest of this section
we will assume that $\det(D) \ne 0$, and hence that no eigenvalue of $\TT$
is exactly equal to one.

We now turn to the construction of the domain-wall propagator
$G_F=D_F^{-1}$. Using \seeq{DDag} one has
\beq
  G_F = D_F^\dagger\, \g_5 B^{-1/2}\, G\, \g_5 B^{-1/2} \:,
\label{GF}
\eeq
where $G=\O^{-1}$. Our task is to find an explicit representation
for $G$. We begin by writing the spectral decomposition 
\beq
  \TT = \sum_i \sket{v_i} \l_i \sbra{v_i} \:.
\label{specTT}
\eeq
For each eigenvalue, we define $q_i=\min(\l_i,\l_i^{-1})$.
We now construct a new matrix
\beq
  Q = \sum_i \sket{v_i} q_i \sbra{v_i} \:,
\label{Q}
\eeq
with the property $0<{\rm spec}(Q)<1$. Next we consider the 
infinite-size matrix $\O_\infty$ constructed from the translationally
invariant part of $\O$, extended to the range $-\infty<s,s'<\infty$.
The inverse $G_\infty=\O_\infty^{-1}$ is 
\bqry
  G_\infty(s,s') & = & \sum_i \sket{v_i} q_i^{|s-s'|} f(q_i) \sbra{v_i} \\
  & = & Q^{|s-s'|} f(Q) \:,
\label{Ginfty}
\eqry
where $f^{-1}(q)=q^{-1}-q>0$. \seEqs{Omega} and\seneq{YTT} imply 
$\sum_s \O_\infty(s'',s)\,G_\infty(s,s')=0$ for $s'' \ne s'$,
while the correct normalization for $s'' = s'$ is ensured by
the presence of $f(Q)$. Since it was constructed using $Q$ (and not $\TT$), 
$G_\infty(s,s')$ vanishes for large $|s-s'|$. 

Returning to the finite-$s$ case, we now have for $1 \le s,s' \le N_s$
\beq
  G(s,s') = G_\infty(s,s') + H_{++}(s,s') + H_{--}(s,s')
                           + H_{+-}(s,s') + H_{-+}(s,s') \:,
\label{G}
\eeq
where
\bqry
  H_{++}(s,s') & = & Q^{s}       \, A_{++} \, Q^{s'}       \NON
  H_{--}(s,s') & = & Q^{N_s+1-s} \, A_{--} \, Q^{N_s+1-s'} \label{H}\\
  H_{+-}(s,s') & = & Q^{s}       \, A_{+-} \, Q^{N_s+1-s'} \NON
  H_{-+}(s,s') & = & Q^{N_s+1-s} \, A_{-+} \, Q^{s'}       \:. \nonumber
\eqry
The four-dimensional matrices $A_{\pm\pm}$ solve a system of linear
equations given in Appendix~A, and have a convergent $N_s\to\infty$ limit.
Hermiticity of $G$ implies $A_{-+}=A_{+-}^\dagger$.

The physical significance of the above construction is the following.
In a fixed gauge-field background $\cu=\{ U_{x\m} \}$ one has
$G_F(s,s';\cu)=\svev{\j(s)\,\bar\j(s')}_\cu$.
Suppose that we project $\bar\j(s')$, 
the anti-fermion field on a given $s'$-slice,
onto the (four-dimensional) state $\g_5 B^{1/2} v_i$
where $v_i$ is an eigenvector of $\TT$. Using \seeqs{GF} and \seneq{G}, 
each term in the corresponding projection 
of $\svev{\j(s)\,\bar\j(s')}_\cu$
must have an $s'$-dependence given by either 
$\l_i^{+ s'}$ or $\l_i^{-s'}$ ($\l_i$ is the corresponding eigenvalue).
The $s'$-dependence will involve no other eigenvalue of $\TT$.
A similar statement can be made for
the $\j$-field. To this end, we write 
$G_F = (D_F^\dagger D_F)^{-1}D_F^\dagger$ and use \seeq{rg5} to related
$(D_F^\dagger D_F)^{-1}$ to $(D_F D_F^\dagger)^{-1}$.
This reproduces a key feature of the second-quantized
transfer matrix formalism~\cite{NNold,VF}, but now in the lagrangian formalism,
which is directly related to the manner one simulates fermionic
correlation functions. 

We end this section with two technical comments. In the second-quantized
transfer matrix formalism one can define $s$-dependent operators via 
$\hat{a}_{L,R}(s) = \hat{T}^s \hat{a}_{L,R} \hat{T}^{-s}$,
where $\hat{T}$ is the second-quantized version of $T$ 
(defined by $\hat{T}=\exp(\hat{a}^\dagger\, \log T \, \hat{a})$,
where $\hat{a}$ and $\hat{a}^\dagger$ are creation and annihilation
anti-commuting operators). 
As noted in ref.~\cite{VF} (see eq.~(A.9) there) 
using the transfer matrix $\hat{T}$ implies that
the operators $\hat{a}_R(s)$ and $\hat{a}_L(s)$ are identified with the
Grassmann variables $\j_R(s)$ and $\j_L(s-1)$ respectively. 
In other words, related Grassmann variables and operators do
not always have the same $s$-coordinate. 
On the other hand, if one uses the second-quantized version
of $\TT$, then related Grassmann variables and
operators do have the same $s$-coordinate in all cases.
This explains the appearance of $\TT$ instead of $T$ in the expressions for 
the domain-wall propagator.

Our last comment concerns the new overlap formalism.
Dropping all terms of order $a_5^2$ and higher in $\TT$ one has
\beq
  \TT \sim 
  \left( \begin{array}{cc}
  1 + a_5 (M-W) &  a_5 C    \\
  a_5 C^\dagger &  1 - a_5 (M-W)
       \end{array}\right) \,.
\label{TTlin}
\eeq
If we now take the limit $a_5 \to 0$ while keeping the product
$a_5 N_s = L_s$ fixed, we have $\TT^{N_s}(a_5) \to \exp(\g_5 D L_s)$.
This reproduces the result of ref.~\cite{KY}.
Recognizing that $\g_5 D$ is the new overlap's hamiltonian, we see
that both the old and the new overlap formulae for QCD can be
recovered as suitable limits of domain-wall fermions.

\vspace{5ex}
\noindent {\large\bf 4.~~Complex transfer matrix}
\vspace{3ex}
\secteq{4}

In this section we discuss what happens if $Ma_5>1$, 
starting with the free-field case. 
Since $[B_0,C_0]=0$, one can lump together $B_0^{\pm 1/2}$ factors,
so that only $B_0^{\pm 1}$ will occur in the equations that define
$\O_0$ and $Y_0$ ({\it cf.} \seeqs{DDag} and\seneq{Y}).
One has 
\beq
  Y_0(p_\m) = B_0^{-1}(p_\m)[1+B_0^2(p_\m)+\sum_\m \sin^2(p_\m)] \:.
\eeq
Note that the sign of $Y_0(p_\m)$ is determined by the sign of $B_0(p_\m)$,
and that either $Y_0(p_\m)>2$ or $Y_0(p_\m)<-2$. 
The eigenvalues $\TT_0(p_\m)$ are all real, and have the same
sign as $B_0(p_\m)$ ({\it cf.} \seeq{YTT}).
Therefore, the free-field propagator exhibits sign oscillations,
but otherwise everything stays pretty much the same as in the $Ma_5<1$ case.

The situation is different in the interacting theory,
where $[B,C] \ne 0$. When $Ma_5>1$, $B$ has both positive and
negative eigenvalues, and $B^{1/2}$ has both real and imaginary eigenvalues. 
Consequently, the transfer matrix $\TT$ is no longer hermitian. 
(\seEq{TKK} still holds 
if the {\it same} definition of $B^{\pm 1/2}$ is used in the expressions for
$K$ and $K^\dagger$, {\it cf.} \seeq{K}, 
even though $B^{1/2}$ is no longer hermitian.
Of course, $K^\dagger$  does {\it not} stand for 
the hermitian conjugate of $K$ in this case.)

While $\det(D_F)$ is always real due to \seeq{rg5}, 
it is not necessarily positive for $Ma_5>1$.
In particular $\det(D_F)$ may occasionally vanish.
(This is not true for $Ma_5<1$ and $m>0$, see Appendix~A.)
The propagator exists except on the measure-zero subset defined
by $\det(D_F) = 0$, and can be constructed
using the same technique as in Sec.~3.
The matrix $Q$ is now defined as follows.
Let $y_i$ denote an (in general complex) eigenvalue of $Y$.
For $y_i$ that does not belong to the closed interval $[-2,2]$ 
on the real axis, we define $q_i$ to be the solution
of $q_i + q_i^{-1} = y_i$ obeying $|q_i|<1$.  
For $y_i \in (-2,2)$ the roots obey $|q_i|=1$ while $q_i \ne q_i^{-1}$.
In this case we arbitrarily pick one of the roots.
Finally, we disregard gauge-field configurations leading to any $y_i=\pm 2$, 
hence $f(q_i)$ exists. (This amounts to ignoring another measure-zero set;
see also the last paragraph of Appendix~A.) 
The rest of the construction is the same as before.

Again, as discussed in Sec.~3, 
the dependence of the propagator on $s$ and $s'$ is 
governed by the (now complex) eigenvalues and eigenvectors of the 
transfer matrix $\TT$.
Hence, slowly decreasing anomalous effects may now
arise from eigenvalues lying anywhere close to the unit circle.

When $Ma_5>1$, both $B^{-1/2}$ and $\TT$ are unbounded.
The possibility that $\TT$ may have very large eigenvalues is, however,
not worrisome. The latter correspond to very small eigenvalues of $Q$,
and so they lead to very short-range correlations in the $s$-direction.

Of special significance is the chiral Ward identity that governs the
pion mass. For any $a_5$ and $M$, 
the anomalous term in the pion-mass Ward identity 
is positive when the number of dynamical flavors is even,
as well as in the quenched case (see Appendix~B of ref.~\cite{VF}).
Thus, the anomalous contributions to this particular Ward identity,
coming from all (real or complex) eigenvalues of the transfer matrix, always
add up.

Because of \seeqs{Y} and\seneq{YTT}, close-to-unity eigenvalues of
the non-hermitian $\TT$ should still correspond to approximate zero modes 
of the hermitian operator $\g_5 D$.
The latter have been extensively studied recently~\cite{PV,SCRI2,COL}.
To date, however, no information exist on the distribution of 
eigenvalue in all the rest of the complex plane.
In particular, it is not known how many eigenvalues are located
near the unit circle {\it away} from the point one on the positive real axis.
Therefore it is also not known how much of the anomalous effect
observed in numerical simulations~[1-4] is due to 
(approximate) zero modes of  $\g_5 D$. 
The above issues clearly deserve a more detailed study.

\vspace{5ex}
\noindent {\large\bf 5.~~A modified pseudo-fermion action}
\vspace{3ex}
\secteq{5}

  In the rest of this paper we impose the condition $Ma_5<1$,
and alongside with it the strict positivity of the transfer matrix.
For simplicity we also restrict the discussion to $0<M<2$.
As discussed in the introduction,
the troublesome eigenvalues of the transfer matrix are now the ones close
to unity. In this section we propose a method to reduce 
the density of close-to-unity eigenvalues in a dynamical simulation
by modifying the {\it Pseudo-fermion} 
(also known as Pauli-Villars) part of the action.

Let us first recall why a domain-wall fermion action
must be accompanied by a pseudo-fermion action. From the point of view
of the gauge field, the domain-wall action introduce $N_s$ ``flavors"
of four-dimensional Dirac fermions. For $0<M<2$, only one Dirac
fermion is light (and is identified with a quark field).
The other $N_s-1$ Dirac fields have $O(1)$ masses.
If their number was kept fixed, we could simply ignore them
in the continuum limit. The chiral limit, however, requires $N_s \to\infty$.
If this limit is taken at fixed value of the bare coupling $g$,
the $N_s$-dependent contribution of the heavy ``flavors"
must be subtracted. 

   One can express the domain-wall fermion determinant as
\beq
  \det(D_F) = \m_{F}^{N_s} \times (\mbox{finite factor}) \:.
\label{bulk}
\eeq
The ``finite factor", which accounts for the quark field, has a convergent 
$N_s\to\infty$ limit. The bulk term $\m_{F}^{N_s}$
is the (undesirable) contribution
of the $O(N_s)$ massive ``flavors". Explicit expressions for both terms
were first derived in the transfer matrix formulation~\cite{NNold,VF},
and more recently by direct manipulations of determinants~\cite{HN10}. 
The latter technique will also be used below. Explicitly, 
\beq
  \m_{F} = \prod_{\l_i>1} \l_i \:,
\label{Fbulk}
\eeq
where the product is over the greater-than-one eigenvalues of $\TT$,
{\it cf.} \seeq{TT}.
As expected, the bulk term is independent of the choice of boundary
conditions in the $s$-direction (\ie it is independent of $m$).
One way to cancel the bulk contribution is to introduce 
a five-dimensional boson field $\f_{xs}$, having the same spin and internal
indices as the domain-wall fermions, on a lattice whose 
fifth coordinate ranges only from 1 to $N = N_s/2$. 
We will refer to $\f_{xs}$ as the pseudo-fermion field.
The pseudo-fermion action is
\beq
  S_{\rm pf}^{\rm unmodified} = \sum_{xs,ys'} 
  \f^\dagger_{xs} (D_{\rm pf} D_{\rm pf}^\dagger)_{xs,ys'} \f_{ys'} \:.
\label{Spf}
\eeq
A common choice is $D_{\rm pf}=D_F(ma_5=1)$.
Excepting $ma_5=-1$, in fact, any $O(1)$ value for $ma_5$ will do.
When integrating over both fermions and pseudo-fermions the bulk terms 
cancel out, leaving a convergent $N_s\to\infty$ result that has the same
long-distance behavior in four dimensions as the finite factor in \seeq{bulk}.
(Another option is to use a first-order pseudo-fermion action.
When the transfer matrix has a gap, the bulk term of the 
first-order action converges faster to the fermionic bulk term~\cite{PV}.
This advantage disappears if the gap is small.)

As mentioned above, for $Ma_5<1$ chiral symmetry violations should arise from
close-to-unity eigenvalues of the transfer matrix.
Actually, in the free-field case the spectrum has a gap.
This follows from \seeqs{Y} and\seneq{YTT}, and the fact that
\beq
  D_0 D_0^\dagger(p_\m) = (M-W_0(p_\m))^2 + \sum_\m \sin^2(p_\m) > 0 \:.
\label{freegap}
\eeq
Since the Brillouin zone is compact, the ($M$-dependent) minimal 
eigenvalue of $D_0 D_0^\dagger$ is strictly positive.
Now, the eigenvalues of $D D^\dagger$ are gauge-invariant 
continuous functions of the link variables.
Since $D_0 D_0^\dagger$ has a gap in the free-field case,
one expects that $D D^\dagger$ too should have a 
(somewhat smaller) gap if the local plaquette action
\beq
  \cp_x = \sum_{\m < \n} \Re \tr (1 - U_{x\m}U_{x+\hat\m,\n}
          U^\dagger_{x+\hat\n,\m}U^\dagger_{x\n}) \:,
\label{cp}
\eeq
is very small everywhere. In other words, 
zero modes of $D D^\dagger$ should arise only if $\cp_x$
exceeds an $O(1)$ constant $c_0(M)>0$ at least for {\it some} lattice sites.
Hence, if one is sufficiently close to the continuum limit and
$M$ is not too close to the critical points 0 or 2, 
these zero modes should be suppressed by the plaquette action.
In the context of domain-wall fermions we can therefore regard these
zero modes as lattice artefacts. (See refs.~\cite{SCRI2,ssrs}
for related work. The existence of a $c_0(M)>0$
can be tested numerically. A proof that $c_0(M)>0$
is desirable, but one should keep in mind that rigorous inequalities may
under-estimate the value of  $c_0(M)$.)

For $\b\sim 6$, one is probably not close enough to the continuum limit
to effectively suppress the zero modes of $D D^\dagger$
by the plaquette action alone.
The above consideration lead us to propose a
modified pseudo-fermion action
\beq
  S_{\rm pf} = \sum_{xs,ys'} 
  \f^\dagger_{xs} (D_{\rm pf} D_{\rm pf}^\dagger)_{xs,ys'} \f_{ys'} 
  + \sum_{xs} \f^\dagger_{xs} (c_1 \cp_x)^n \f_{xs} \:,
\label{Smodf}
\eeq
where $c_1$ is a continuous parameter and $n$ is a positive integer.
Assuming $c_0(M)$ is known, a reasonable choice 
is $c_1 \sim c_0^{-1}(M)$. The choice of $n$ is discussed later.
The idea behind the modified pseudo-fermion action is to suppress
in a selective way those gauge-field configurations supporting 
close-to-unity eigenvalues. The aim is to achieve this suppression
in (dynamical-fermion) simulations with not too large $N_s$
and at presently accessible values of $\b$.
As explained below, the main side-effect of the modified action
is expected to be a small renormalization of the coupling constant.

As in \seeq{DDag}, the pseudo-fermion matrix defined by 
\seeq{Smodf} can be written as 
$\g_5 B^{1/2}\, \O_{\rm pf}\, \g_5 B^{1/2}$.
The explicit expression for $\O_{\rm pf}$ is the same as
the \rhs of \seeq{Omega} except for the replacement 
\beq
  Y \to Y_{\rm pf}(c_1) = Y + B^{-1/2}\, (c_1 \cp)^n\, B^{-1/2} \:,
\label{Ypf}
\eeq
where $\cp$ stands for the diagonal matrix $\d_{xy} \cp_x$.
(Similar replacements are made in the definitions of $X_{++}$ and $X_{--}$.
For $ma_5=1$ one has $Y_{\rm pf}=X_{++}^{\rm pf}=X_{--}^{\rm pf}$.)

An expression for $\det(\O_{\rm pf})$ can be written for even $N$
using the general formulae of ref.~\cite{HN10},
see Appendix~B below. The result is
\beq
  \det(\O_{\rm pf}) = \det\!\left(
  R^{-} - P\, T_{\rm pf}^N\, P^{-1}\, R^{+}
  \right) \:,
\label{Opf}
\eeq
where
\bqry
  R^{-} & = & \left(\begin{array}{cc}
  1 & 0 \\ Y_{\rm pf} - X_{--}^{\rm pf} & - X_{+-}
  \end{array}\right)
\label{R-} \\
  R^{+} & = & \left(\begin{array}{cc}
  - X_{-+} & Y_{\rm pf} - X_{++}^{\rm pf} \\ 0 & 1
  \end{array}\right)
\label{R+} \\
  T_{\rm pf} & = & 
  \left(\begin{array}{cc}
  Q_{\rm pf} & 0 \\ 0 & Q^{-1}_{\rm pf}
  \end{array}\right) 
\label{T2}  
\eqry
\beq
  P = 
  \left(\begin{array}{cc}
  1 & Q_{\rm pf} \\ Q_{\rm pf} & 1
  \end{array}\right)\:,
\qquad
  P^{-1} = (1-Q_{\rm pf}^2)^{-1}
  \left(\begin{array}{cc}
  1 & -Q_{\rm pf} \\ -Q_{\rm pf} & 1
  \end{array}\right)  \:.
\label{P}
\eeq
As before, the matrix $Q_{\rm pf}$ is defined by
$Q_{\rm pf} + Q^{-1}_{\rm pf} = Y_{\rm pf}$ and the condition
$0<{\rm spec}(Q_{\rm pf})<1$. One has
\beq
  P\, T_{\rm pf}^2\, P^{-1} = 
  \left(\begin{array}{cc}
  -1 & Y_{\rm pf} \\ -Y_{\rm pf} & Y^2_{\rm pf} -1
  \end{array}\right) \:.
\label{PY}
\eeq
Note that the \rhs of the last equation is a function of $Y_{\rm pf}$ only.

For large $N$, $Q_{\rm pf}^N$ vanishes
whereas $Q_{\rm pf}^{-N}$ grows unboundedly.
Like \seeq{bulk}, one can express $\det(\O_{\rm pf})$ 
as $\m_{\rm pf}^N(c_1)$ times a finite factor, where
\beq
  \m_{\rm pf}(c_1) = \det\!\left( Q^{-1}_{\rm pf}(c_1) \right) \:.
\label{pfbulk}
\eeq
(See \seeq{Ofinite} for the finite factor.)
For $N_s=2N$, the total bulk factor coming from the integration 
over both fermions and pseudo-fermions is 
\beq
  \left( {\m^2_{F} \over \m_{\rm pf}(c_1)} \right)^N \:.
\label{total}
\eeq
In the unmodified case, $c_1=0$, using \seeq{Fbulk} and $\det(\TT)=1$ one has
$\m_{\rm pf}(0) = \m_{F}^2$,
showing that the bulk factors indeed cancel each other.

We now discuss how the bulk factor is modified for
$c_1 > 0$. Since the plaquette term is positive, the $c_1$-derivative
of the eigenvalues of $Y_{\rm pf}$ is always positive,
and the $c_1$-derivative of the eigenvalues of $Q_{\rm pf}$ is always negative.
Therefore $\m^2_{F}/\m_{\rm pf}(c_1)$ is a decreasing function of $c_1$.

Let us now assume that $\TT$ (or $Q$) has an eigenvalue very close to one.
For $M \sim 1.7$, the support of the corresponding eigenvector 
should consist of very few lattice sites~\cite{SCRI2}.
Moreover, as discussed earlier $\cp_x$
is likely $O(1)$ on those sites. As $c_1$ is varied from zero to its
chosen value, we thus expect an $O(1)$ change in the corresponding 
eigenvalue of $Q_{\rm pf}(c_1)$. We will argue below that there should be
only a small change in the product of all eigenvalues which are {\it not} 
close to unity. Consequently, the $O(1)$ change in what used to be 
a close-to-unity eigenvalue implies an $O(1)$ reduction
in the magnitude of $\m^2_{F}/\m_{\rm pf}(c_1)$. This, in turn,
should suppress the Boltzmann weight of the corresponding
gauge-field configuration already for modest values of $N_s$ 
({\it cf.} \seeq{total}).

Eigenvalues of the transfer matrix not too close to
unity are typically not localized, and it is plausible that
the effect of the modified action on them can be accounted for
by perturbation theory. It is easy to see that the leading
perturbative effect of the modified action is to renormalize 
the coupling constant as
\beq
  {1\over g^2} \to {1\over g^2} 
  + n c_1^n \sum_s \svev{\cp_x^{n-1}\,\f^\dagger_{xs}\f_{xs}} \:.
\label{pert}
\eeq
Since the free transfer matrix has a gap, the perturbative value 
of $\svev{\f^\dagger_{xs}\f_{xs}}$ at any finite order, 
and in any gauge-field background, 
is regular. (This is true even if the associated exact
transfer matrix has a unit eigenvalue.
In other words, a singularity in the propagator cannot develop if
we sum the Born series only up to a finite order.) Now, 
the expectation value on the \rhs of \seeq{pert} involves $n$ or more loops. 
Therefore the resulting change in $1/g^2$ should be of order $N_s\, \a^n$. 
Since in practice $\a \sim 0.1$, if we take for example $n=4$
this effect may be at the level of 1\% or less,
when $N_s$ is in the range of 10 to 100.
Thus, the leading side-effect of the modified action 
seems to be an innocuous, relatively small change in the bare coupling.

In short, we believe that perturbation
theory can be trusted for the collective contribution of all eigenvalues,
{\it except} when there are close-to-unity ones. 
Since the perturbative effect should be small, 
an $O(1)$ change in $\m_{\rm pf}(c_1)$ should
take place only when there are close-to-unity eigenvalues,
and this change works in the direction of suppressing the Boltzmann
weight of the corresponding gauge-field configurations. 

The above arguments, while plausible, are heuristic.  
One question that can be settled by an explicit (perturbative) calculation
is whether the loop integrals in $\svev{\cp_x^{n-1}\,\f^\dagger_{xs}\f_{xs}}$,
while regular, happen to produce large numerical factors. 
A calculation of this expectation value is also necessary in
order to be able to compare results at different values of $c_1$,
while maintaining a fixed value of the effective bare coupling 
({\it cf.} the \rhs of \seeq{pert}).

Last we discuss how the $N_s\to\infty$ limit may be taken with the modified
action. Evidently, if all other parameter were kept fixed,
then in the limit  $N_s\to\infty$ the perturbative effects induced
by the modified pseudo-fermion action would eventually run out of control.
While the strict  $N_s\to\infty$ limit is not very useful for
practical purposes, it is legitimate to ask whether, in principle, the modified
action has a sensible $N_s\to\infty$ limit.

If we allow both $N_s$ and $n$ to grow, the limit may
in fact depend on the ratio of these two numbers.
As an example, one option is to take the limit $n\to\infty$ {\it before} 
the limit $N_s\to\infty$. Sending $n$ to infinity has the following
effect. For $c_1 \cp_x < 1$, the limit of $(c_1 \cp_x)^n$ is zero. 
Hence, the modification vanishes if $c_1 \cp_x < 1$ for all $x$. 
If, on the other hand, $c_1 \cp_x > 1$ even for a single site, 
the norm of $(c_1 \cp_x)^n$ will blow up,
and along with it some of the eigenvalues of $Q_{\rm pf}^{-1}$.
In summary, in the limit $n\to\infty$ the bulk factor
$\m^2_{F}/\m_{\rm pf}(c_1)$ is unchanged
if $c_1 \cp_x < 1$ for all $x$, whereas it vanishes
if $c_1 \cp_x > 1$ for any lattice site. Therefore, 
the  limit $n\to\infty$ amounts to imposing the constraint 
$c_1 \cp_x < 1$ on the gauge-field configuration space.
While this constraint should not change the continuum limit,
throwing out all configurations with some $c_1 \cp_x > 1$ 
in a completely unselective manner could slow down simulations. 
As argued above, for moderate values of $N_s$ and relatively small values
of $n$, the modified action may do a good job in suppressing
only those gauge-field configurations supporting close-to-unity
eigenvalues, leading to a minimal ``waste" of configurations.

\vspace{5ex}
\noindent {\large\bf 6.~~Conclusions}
\vspace{3ex}
\secteq{6}

In this paper we have studied potential sources for the anomalous
term in chiral Ward identities.
For conventional domain-wall fermions ($a_5=1$)
the transfer matrix is complex if $M>1$.
As discussed in Sec.~4, in the numerical work of refs.~[1-4]
slowly-suppressed chiral symmetry violations could therefore arise 
not only from eigenvalues which are close to one, but in fact
from eigenvalues in the vicinity of the entire unit circle.

For $a_5=0.5$ and $0<M<2$ the transfer matrix will be strictly positive.
The first question that has to be addressed numerically is whether
the favorable range for a single light quark is still $M \sim 1.7$,
as in the $a_5=1$ case~\cite{BS,COL}. The range $M \sim 1.7$ seems reasonable
also from the point of view of the new overlap formulation~\cite{SCRI2}.
Since the new overlap involves a continuous $s$-coordinate limit,
it is plausible that the best value of $M$ may be rather insensitive
to $a_5$.

For fixed $1<M<2$, as $a_5$ is decreased from 1 to 0.5
all eigenvalues must flow towards the positive real axis. 
Depending on the flow pattern, the number of close-to-one eigenvalues 
at $a_5=0.5$ could be quite different
from the original number of eigenvalues close to the unit circle.
Hence, the transition from $a_5=1$ to $a_5=0.5$ could by itself have a 
significant effect on the anomalous term. 
Finally, once the above issues are resolved,  
one can proceed to test whether the modified pseudo-fermion action 
is useful in further reducing lattice-artefact
violations of chiral symmetries at presently accessible values of $N_s$.

\vspace{5ex}
\noindent {\bf Acknowledgements}
\vspace{3ex}

I thank the participants of the
RIKEN BNL Research Center workshop on {\it Fermion Frontiers in
Vector Lattice Gauge Theories}, held on May 6-9, 1998, 
at Brookhaven National Laboratory,
for extensive discussions that motivated this work.
Especially I would like to thank the organizers for creating this
unique opportunity to exchange ideas on the subject.
This work is supported in part by the Israel Science Foundation.

\newpage
\vspace{5ex}
\noindent {\bf Appendix A. The second-order propagator}
\vspace{3ex}
\secteq{A}

\noindent For  $2 \le s'' \le N_s-1$, \seeqs{Omega} and\seneq{YTT} imply 
\beq
  \sum_{s} \O(s'',s)\, G_\infty(s,s') -\d_{s'',s'} = 
  \sum_{s} \O(s'',s)\, H_{\pm\pm}(s,s') = 0 \:.
\eeq
For $s''=1$ and $s''=N_s$ there are boundary effects.
The above expressions are not zero, and their $s'$-dependence
is given by $Q^{\pm s'}$. The linear combination that gives the propagator
$G$ (\seeq{G}) is determined by requiring the coefficients of $Q^{\pm s'}$
to vanish. By imposing this condition on the $Q^{+s'}$ part we obtain
\beq
  {\cal C} \left( \begin{array}{c} A_{++} 
                                \\ A_{-+} \end{array} \right) =
  \left( \begin{array}{c} Q^{-2} - X_{++}Q^{-1} 
                       \\ - X_{+-}Q^{-1} \end{array} \right) f(Q)\,,
\eeq
and by imposing this condition on the $Q^{-s'}$ part we obtain
\beq
  {\cal C} \left( \begin{array}{c} A_{+-} 
                                \\ A_{--} \end{array} \right) =
  \left( \begin{array}{c} - X_{-+}Q^{-1}
                       \\ Q^{-2} - X_{--}Q^{-1}  \end{array} \right) f(Q)\,,
\eeq
where
\beq
  {\cal C} =  \left( \begin{array}{cc}  
     X_{++}Q - Q^2 + X_{-+}Q^{N_s}  &  X_{++}Q^{N_s} - Q^{N_s-1} + X_{-+}Q \\
     X_{--}Q^{N_s} - Q^{N_s-1} + X_{+-}Q  &  X_{--}Q - Q^2 + X_{+-}Q^{N_s}
              \end{array} \right) \:.
\eeq
Note the ${\cal C}$ has a convergent $N_s\to\infty$ limit. A second-quantized
transfer-matrix representation of the finite-$N_s$ fermion determinant 
is given in ref.~\cite{VF} (see in particular eqs.~(3.1) and (3.10) therein).
We decude from it that for $Ma_5<1$ and $m>0$, 
$\det(D_F)$ is (real and) strictly positive for any number of flavors. 
Therefore $D_F$ has an inverse, and the above equations must
have a solution which obeys $A_{-+}=A_{+-}^\dagger$.

We note that $D_F^{-1}$ may exist even if $\TT$ has an eigenvalue which
is exactly equal to one. In this case,
a possible way to construct $D_F^{-1}$ is to perturb
the background field so that that eigenvalue will be only approximately
equal to one, say to up $O(\e)$. The construction of Sec.~3 is now applicable.
$D_F^{-1}$ for the initial background field can then be found by
carefully removing the perturbation, keeping track of the leading
terms in $\e$.

\vspace{5ex}
\noindent {\bf Appendix B. The second-order determinant}
\vspace{3ex}
\secteq{B}

For even $N$,
we can use the formulae for the determinant of a general tridiagonal matrix,
derived in the appendix of ref.~\cite{HN10},
to write down an expression for $\det(\O_{\rm pf}(c_1))$, {\it cf.} \seeq{Opf}.
(Here the block entries of the tridiagonal matrix have spin indices ranging 
from one to four. With the replacement $N \to N_s$, a similar formula holds 
for $\det(\O)$, {\it cf.} \seeq{Omega}.)
In the notation of ref.~\cite{HN10}, the present application is defined by
\bqry
  \a & = & 
  \left(\begin{array}{cc}
  -1 & 0 \\ Y_{\rm pf} & -1
  \end{array}\right) 
  = \a_j \:, \qquad j=1,\ldots,N/2-1 \:,
\\
   \a_{N/2} & = & 
  \left(\begin{array}{cc}
  -1 & 0 \\ X_{--}^{\rm pf} & X_{+-}^{\rm pf}
  \end{array}\right)  \:,
\\
  \b & = & 
  \left(\begin{array}{cc}
  -1 & Y_{\rm pf} \\ 0 & -1
  \end{array}\right) 
  = \b_j \:, \qquad j=1,\ldots,N/2-1 \:,
\\
   \b_{N/2} & = & 
  \left(\begin{array}{cc}
  X_{-+}^{\rm pf} & X_{++}^{\rm pf} \\ 0 & -1
  \end{array}\right)  \:.
\eqry
With these definitions, eq.~(A.10) of ref.~\cite{HN10} reads
\beq
  \det(\O_{\rm pf}) = \det\!\left[
  \a^{-1}\,\a_{N/2} - \Big(-\a^{-1}\b\Big)^{N/2}\, \b^{-1}\,\b_{N/2}\right] \:.
\eeq
Substituting the above explicit expressions we arrive at \seeq{Opf}.
Note that $(-\a^{-1}\b\Big)$ is equal to the \rhs of \seeq{PY}.
The bulk factor of $\det(\O_{\rm pf})$
can be separated as follows. We first rewrite \seeq{Opf} as 
\beq
  \det(\O_{\rm pf}) = \det\!\left(1-Q^2_{\rm pf}\right)
  \det\!\left(
  S^{-} - T_{\rm pf}^N\, S^{+}
  \right) \:,
\label{Opfs}
\eeq
where $S^{\pm} = P^{-1} R^{\pm}$. Using \seeq{T2} it is now easy to check that
\bqry
  \det(\O_{\rm pf}) & = &  
  \det\!\left( Q^{-N}_{\rm pf} \right) \:
  \det\!\left(1-Q^2_{\rm pf}\right) \NON
  & & \times \:
  \det\!\left[
     \left(\begin{array}{cc}
     1 & 0 \\ 0 & Q^N_{\rm pf}
     \end{array}\right) S^- - 
     \left(\begin{array}{cc}
     Q^N_{\rm pf} & 0 \\ 0 & 1
     \end{array}\right) S^+ 
  \right] \:.
\label{Ofinite}
\eqry
The bulk factor $\det\!\left( Q^{-N}_{\rm pf} \right)$ appears explicitly in
the above equation. The other terms are by definition the finite factor. 
For $Ma_5<1$ the entries of $\O_{\rm pf}$ are bounded, 
and $\det(\O_{\rm pf})$ is finite for any finite $N$.
This is more easily seen using \seeqs{Opf} and\seneq{PY}.
It can be checked that if an eigenvalue of $Q$ approaches one,
then the restriction of the \rhs of \seeq{PY} to this eigenvalue 
has a finite limit. Moreover, the restriction of
$P\, T_{\rm pf}^N\, P^{-1}$ to this eigenvalue 
grows only linearly (not exponentially) with $N$.


\vspace{5ex}
\centerline{\rule{5cm}{.3mm}}

\end{document}